\documentstyle[preprint,aps,prb]{revtex}

\begin{document}


\title{Noise in single quantum well infrared photodetectors}

\author{M. Ershov}
\address{Department of Computer Software, University of Aizu,
  Aizu-Wakamatsu 965-80, Japan}  
\author{A. N. Korotkov} 
\address{Nuclear Physics Institute, Moscow State University,
        Moscow 119899, Russia, \\
and Fundamental Research Laboratories, NEC Corporation,
Tsukuba, Ibaraki 305, Japan}

\date{\today} 
\maketitle

\begin{abstract}
The spectral density of current fluctuations in single quantum
well infrared photodetectors is calculated using Langevin
approach. The noise gain and the photocurrent gain are expressed in
terms of basic transport parameters. Fluctuations of the incident
photon flux are taken into account.

\end{abstract}

\narrowtext
\vspace{1ex}

The noise of the dark current and photocurrent is an important factor
in operation of the Quantum Well Infrared Photodetectors
(QWIPs).\cite{LevineR} For typical operating conditions, the main
source of fluctuations in QWIPs is generation-recombination noise
associated with the excitation of carriers from the QWs into continuum
and their capture into the QWs.  The study of noise in QWIPs is
important for applications to obtain the devices with high
detectivity, and it also provides an additional physical insight for 
these systems.  The fluctuations are
the natural sources of the transient excitation, therefore the QWIP
response to this excitation provides the information on internal
physical processes, and thus forms the basis for the noise
spectroscopy.\cite{ZielBFl59,KoganNB96} Due to the numerous
experimental and theoretical works on noise properties of QWIPs with
multiple QWs (see, e.g., Refs.\
\cite{LevineR,LiuNAPL92,BeckN93,LiuN94,WangN94,ChoiAPL94,ShadrinN1}),
the overall understanding of the QWIP noise characteristics is
satisfactory. Some issues, however, are still controversial, such as
the relation between the noise gain and the  photocurrent (optical)
gain, the frequency dispersion of the noise spectral density, and the
role of the injecting contact.

In this letter we consider QWIP with single QW (SQWIP).  SQWIP is
especially attractive theoretically because its simple structure
allows an accurate self-consistent calculation of the electric field,
injection current, and charge accumulation in the QW, and hence
a better understanding of the noise properties.  In addition,
SQWIPs are intrinsically fast devices promising for CO$_2$-laser based
high-speed applications, for which the noise is an important
consideration.\cite{LiuFast86,KingstonB79,Liu82GHz} Although SQWIPs
have been studied by several research groups (see, e.g., Refs. 
\cite{LiuSSST91,RosencherS92,BandaraS93,BandaraST93}), we are
aware of only one theoretical paper \cite{CarboneS97} dealing with the
noise in SQWIPs. However, the result obtained in
Ref.\ \cite{CarboneS97} requires a number of restrictive
assumptions (large photocurrent gain, absence of the electron
transport from the QW to emitter, etc.).

In this letter we extend recent theoretical studies of
SQWIPs \cite{Ryzhii_S_JAP95,Ryzhii_S_SST95,ErshovIEEE96} to present a
noise theory for SQWIPs. The frequency-dependent spectral density of
current fluctuations is expressed in terms of the basic transport and
injection parameters, and is applicable for SQWIPs with different
design concepts.

The SQWIP under consideration contains a single-level QW separated by
undoped barriers from heavily doped contacts (Fig.\ 1).
Our model basically follows
Refs.\ \cite{Ryzhii_S_JAP95} and \cite{ErshovIEEE96},
however, we do not assume any particular shape of the emitter and
collector barriers, so for example the injection current $I_e$ has a
general dependence on the electric field $E_e$ in the emitter barrier.
An important transport parameter of the model is the efficiency
$\beta$ ($\beta \le 1$) being the probability for an injected electron
to pass from emitter directly to collector, while $(1-\beta)$ is the
probability of electron capture by QW (here the meaning of the QW capture
probability is different from that in the case of drift electron
transport in QWIPs with multiple QWs -- see Ref.\ \cite{ShadrinN1}).  
The electrons emitted from the QW are
collected by the collector and emitter with probabilities $\zeta$ and
$(1-\zeta)$, respectively (electron transport from the QW to emitter
is especially important for SQWIPs with triangular 
barriers \cite{Ryzhii_S_JAP95,Ryzhii_S_SST95,ErshovIEEE96}). 
Electron transport across the emitter and
collector barriers is assumed to be instantaneous, therefore, our
analysis is limited by frequencies $\omega \ll v_T/\max (W_e,W_c)\sim
10^{12}$ s$^{-1}$, where $v_T$ is a typical (thermal) velocity and
$W_e$, $W_c$ are the emitter and collector barrier thicknesses,
respectively (the model of instantaneous jumps can be used for both
thermoactivated emission and tunneling).  We also neglect the electron
interactions during traveling in the barrier regions, and
single-electron correlations.  The dynamics of the electron transport
in SQWIP is described\cite{notation} by the following Langevin equations:
        \begin{eqnarray}
I=\frac{W_e}{W} \left( I_{ew}-I^t_{we}-I^p_{we}+
\xi_{ew}-\xi^t_{we}-\xi^p_{we} \right) 
        \nonumber \\
+ \, \frac{W_c}{W} \left( I^t_{wc}+I^p_{wc}
+\xi^t_{wc}+\xi^p_{wc} \right) + I_{ec}+\xi_{ec} \, ,
        \label{current}\end{eqnarray}
        \begin{eqnarray}
{\dot Q} =  I_{ew}-I^t_{we}-I^p_{we} - I^t_{wc} - I^p_{wc} \,\,
        \nonumber \\ 
+ \, \xi_{ew}-\xi^t_{we}-\xi^p_{we} -\xi^t_{wc}-\xi^p_{wc} \, .
        \label{Q}\end{eqnarray} 

        \noindent  Here $I$ is the total current through SQWIP 
(which is equal to the
sum of the conduction and displacement currents at any cross section),
$I_{ec}=\beta\, I_{e}$ and $I_{ew}=(1-\beta )\,I_e$ are the currents
from the emitter to collector and well, respectively, and
$I_{wc}=\zeta\, I_w$ and $I_{we}=(1-\zeta)\, I_w$ describe the electron
transport from the well to collector and emitter ($I_w=I_w^p+I_w^t$
where superscripts $t$ and $p$ denote the currents due to thermo- and 
photoexcitation, respectively). We neglect the current from the collector
assuming sufficiently large bias voltage $V$. 
The well is assumed
to be narrow, so the total structure thickness is $W=W_e+W_c$, while
the finite well thickness in the first approximation can be taken into
account via effective values for $W_e$ and $W_c$.
 The current $I_e$ 
depends mainly on the electric field $E_e$ in the emitter,
$E_e=E^0_e +V/W-QW_c/(\varepsilon\varepsilon_0 A W)$,  
while $I_w$
depends on the electron population in QW and electric fields in the
barriers, \cite{RosencherS92,BandaraS93,Ryzhii_S_JAP95} so all
currents are some functions of the QW charge $Q$ (here $A$ is the SQWIP
area, $\varepsilon\varepsilon_0$ is the dielectric constant, and $E^0_e$ 
is the parameter of SQWIP design).   For simplicity we
neglect the dependencies of $\beta$ and $\zeta$ on the accumulated
charge.

Studies of the steady-state characteristics and admittance of SQWIPs
described by similar models have been reported
recently. \cite{RosencherS92,BandaraS93,Ryzhii_S_JAP95,ErshovIEEE96} 
In this letter we concentrate on the fluctuations which in 
Eqs.\ (\ref{current})--(\ref{Q}) are caused by random Langevin 
terms $\xi (t)$.
        All random terms (except $\xi_{we}^p$ and $\xi_{wc}^p$ --
see below) have no mutual correlation, all of them are $\delta$-correlated
in time, and the corresponding spectral densities $S(\omega )$ are given 
by usual Schottky formula:
\begin{equation}
S_{\xi_{ew}} (\omega ) =2\, e\, \langle I_{ew}\rangle , \,\, 
S_{\xi_{wc}^t} (\omega )=2\,e\, \langle I_{wc}^t \rangle , \,\, \mbox{etc.}  
\label{xi}\end{equation}
(brackets denote time averaging). The fluctuations of the photocurrent
depend on the photon source noise. Let the photon flux incident to the
QW has the spectral density $S_p (\omega ) =2\nu [1+\alpha (\omega )]$
where $\nu$ is the average flux (photons per second) and $\alpha$
describes the deviation from Schottky level. Then the noise of the
photoexcitation currents from the QW is given by
\begin{eqnarray}
S_{\xi_{wc}^p} (\omega ) = 2\,e\, \zeta\, \langle I_w^p \rangle 
[1+\zeta\, \eta\, \alpha (\omega )] , 
\,\, \langle I_w^p \rangle = e\,\eta\,\nu ,
        \nonumber \\
S_{\xi_{we}^p} (\omega) = 2\,e\, (1-\zeta ) \langle I_w^p \rangle 
[1+(1-\zeta)\, \eta\,\alpha  (\omega )] , 
        \nonumber \\
S_{\xi_{we}^p\xi_{wc}^p} (\omega) = 2\,e\, \langle I_w^p \rangle 
\, \zeta\, (1-\zeta)\, \eta\, \alpha  (\omega )  
        \label{xi-p}
\end{eqnarray}

\noindent Here the last equation describes the mutual spectral density, 
and the absorption quantum efficiency $\eta$ includes the finite
probability for a photoexcited electron to escape from the QW.
   Equations (\ref{xi-p}) can be derived separating in the 
correlation functions the terms corresponding to one and two  
excitation events. The first term is proportional to the probability
$\xi \eta$ or $(1-\xi )\eta$ while the second term is proportional to
the corresponding product of probabilities. Equations (\ref{xi-p})
show that the
photoexcitation current noise has the simple Schottky behavior only
if the photon flux is Poissonian ($\alpha =0$) or $\eta$ is small.

Equations (\ref{current})--(\ref{xi-p}) allow us to calculate the
noise properties of SQWIP.  Applying the standard Langevin
method \cite{ZielBFl59,KoganNB96} to the linearized version of Eqs.\
(\ref{current})--(\ref{Q}), we first formally solve Eq.\ (\ref{Q}) in the
frequency representation taking into account the dependence of
currents on the accumulated charge. Substituting the result into Eq.\
(\ref{current}) and using Eqs.\ (\ref{xi})--(\ref{xi-p}) for Langevin
sources we obtain the following spectral density of the total current
\begin{eqnarray}
S_I(\omega )= 2e \langle I_{ec} \rangle 
\, + \, 2 e [ \langle I_{ew}\rangle + \langle I_{we}^t \rangle
        \nonumber \\
   +  \langle I_{we}^p \rangle (1+(1-\zeta)\eta\alpha (\omega )) ]
        \left| \frac{W_e}{W} -\frac{\chi}{1-i\omega \tau} \right|^2
        \nonumber \\  
  +2 e \left[  \langle I_{wc}^t \rangle +
     \langle I_{wc}^p \rangle (1+\zeta \eta\alpha (\omega)) \right]
        \left|\frac{W_c}{W} +\frac{\chi}{1-i\omega \tau} \right|^2 
        \nonumber \\
  +4e \left( \langle I_{we}^p \rangle  +\langle I_{wc}^p \rangle \right)
    \, \zeta (1-\zeta) \, \eta\alpha (\omega ) 
\,\,\,\,\,\,\,\,\,\,\,\,\,\,\,\,\,\,\,\,
        \nonumber \\
        \times \mbox{Re} 
     \left[\left( \frac{W_e}{W} -\frac{\chi}{1-i\omega \tau}\right) 
     \left( \frac{W_c}{W} +\frac{\chi}{1+i\omega \tau } \right)\right],    
        \label{noise}\end{eqnarray}
        \begin{equation}
\tau^{-1} = -\frac{dI_{ew}}{dQ} +\frac{dI_{we}}{dQ}+ \frac{dI_{wc}}{dQ} \, ,
        \label{tau}\end{equation} 
        \begin{equation}
\chi =\tau 
\left[ - \frac{dI_{ec}}{dQ}
 + \frac{W_e}{W} \left( \frac{dI_{we}}{dQ} -\frac{dI_{ew}}{dQ}  \right) 
- \frac{W_c}{W}\frac{dI_{wc}}{dQ}  \right] .
        \label{chi}
\end{equation}

\noindent   Here the derivatives $dI_i/dQ$ take also into account the
dependence via the electric field modulated by $Q$.
(Obviously $\langle I_{ew} \rangle = \langle I_{we} \rangle
+\langle I_{wc}\rangle$.)   One can see that
in the case $\alpha (\omega )=const$ the spectral density has a
Lorentzian shape (with pedestal) with the characteristic frequency
$\tau^{-1}$. Equation (\ref{noise}) can be used to determine the noise
gain which is traditionally defined as $g_n(\omega )\equiv S_I(\omega
)/4e\langle I\rangle $.

Equations (\ref{current})--(\ref{Q}) without noise
sources can be also used to calculate the photocurrent gain (the ratio 
between the variations of the total current and the photoexcitation 
current for small-signal harmonic infrared excitation):
\begin{equation}
g_p(\omega ) \equiv \frac{ \delta I(\omega )}{\delta I_w^p (\omega )} = 
\zeta - \frac{W_e}{W} + \frac{ \chi}{1-i\omega \tau}.  
        \label{photogain}
\end{equation}
The frequency dependence of the photocurrent gain is obviously 
governed by the same time constant $\tau$ of the QW 
recharging. \cite{ErshovIEEE96} 
(This time constant corresponds to the characteristic
time of establishing equilibrium at the injecting contact in QWIPs
with multiple QWs. \cite{ErshovAPLTr})

To simplify the further analysis let us assume  
 $|dI_{we}/dQ+dI_{wc}/dQ| \ll
|dI_{ew}/dQ|$ (that is a typical experimental case).
Then $\tau =(1-\beta)^{-1}(-dI_e/dQ)^{-1}$   
and $\chi = \beta /(1-\beta )+W_e/W$. If we also assume 
$\alpha (\omega ) I_w^p \ll I_w$ (so that we can
neglect the non-Poissonian term of the photocurrent), then 
the noise gain is given by
\begin{eqnarray}
   g_n(\omega )=  g_n(\infty )+\frac{g_n(0)-g_n(\infty )}
   {1+(\omega \tau )^2}, \,\, 
   g_n(0)=  \frac{1}{2} \,  \frac{1+\beta}{1-\beta}\, , 
        \nonumber \\
   g_n(\infty )=  \frac{1}{2} + (1-\beta )\, \frac{W_e}{W}\,
   \frac{W_e/W-\zeta} {\beta +\zeta -\beta\zeta} \, .
\label{n-gain}
\end{eqnarray}

Under the same assumptions  the ratio between the noise gain and 
photocurrent gain at small frequencies is given by the expression 
\begin{equation} 
g_n(0)/g_p(0)=(1+\beta )/[2(\beta + \zeta -\beta\zeta)]
\label{noise-photo}
\end{equation}

\noindent  (in conventional photoconductors this ratio is close to
unity \cite{KingstonB79} while in our case unity is realized only 
if $\beta \rightarrow 1$ or $\zeta=1/2$), and the minimal 
detectable photon flux $\nu_{min}$ at low frequency is given by
\begin{equation}
\nu_{min} \equiv \frac{\sqrt{S_I(0)\, \Delta f}}{e \eta\, g_p(0)} =
\frac{\sqrt{2eI\, \Delta f}}{e\eta} \frac{\sqrt{1-\beta^2}}{\beta+\zeta
-\beta\zeta} \, ,
\label{nu-min}
\end{equation}
where $\Delta f$ is the bandwidth.

The time constant $\tau \simeq [ (1-\beta) \, W_c/(\varepsilon
\varepsilon_0 W) \times d(I_e/A)/dE_e ]^{-1}$  
for typical SQWIP structures and operating
conditions \cite{ErshovIEEE96}
 can be within a quite wide range ($\sim
10^{-9}-10^{-3}$ s) and depends strongly on the applied voltage, 
temperature, and SQWIP design. 
Because of strong (typically exponential) dependence of $I_e$
on $E_e$, $\tau$ starts to decrease with illumination intensity when
illumination changes $E_e$ considerably (crudely this occurs when the 
photocurrent becomes comparable or larger than the dark current). 
 The dependence of $\tau$ on temperature is
typically exponential,\cite{ErshovIEEE96} $\tau \propto   
\exp(-kT/\varepsilon_a)$, 
where $\varepsilon_a$ is the activation
energy, which can be used for evaluation of the QW parameters from the
measurements of the SQWIP noise or photocurrent characteristics at
different frequencies and temperatures.

At high frequencies, $\omega\gg 1/\tau$, the QW recharging processes are
``frozen'', and the spectral density of current fluctuations is
determined by the shot noise of elementary currents with
appropriate geometrical factors. The photocurrent is due to the
electrons emitted from the QW only,
i.e. the primary photocurrent. \cite{ErshovAPLTr} It is interesting
to note that in the case $\zeta<W_e/W$ the high-frequency photocurrent gain
$g_p(\infty)$ is negative. 

At low frequencies, $\omega\ll 1/\tau$, the SQWIP operates in the
quasistatic regime. The QW charge responds to the external
excitation and modulates the injection current. In this regime the
modulation effect results in a strong enhancement of both the 
photocurrent gain and the noise gain provided $1-\beta \ll 1$. 

If all the electrons injected from the emitter are captured by
the QW ($\beta=0$), then (see Eq.\ (\ref{n-gain})) the low-frequency
noise gain $g_n(0)=1/2$ corresponds to the usual Schottky level. This 
has been observed experimentally in the SQWIP with thin emitter
barrier. \cite{BandaraST93}

In the special case when $\zeta=1$ and $\alpha =0$ 
our main result given by Eq.\ (\ref{noise}) can be compared with the
result of Ref.\ \cite{CarboneS97}. They coincide in the limit of 
high photocurrent gain, however, they are different for finite 
photocurrent  gain because the result of Ref.\ \cite{CarboneS97} 
is not applicable in this case.
We have checked that for $\zeta=1$, $\alpha =0$ the correct expression 
can be obtained also using the Fokker-Plank technique \cite{KoganNB96} 
(averaging 
$\exp (i\omega (t_m-t_n))$ where $t_m$ and $t_n$ are the moments of 
electron jumps). In the general case considered in the present
letter, the Fokker-Plank technique becomes much more cumbersome
than the Langevin method.

In conclusion, we have calculated the noise in the SQWIP under
the assumption of fast electron jumps over (or through) the barriers.

We thank P.~Mazzetti and H.~C.~Liu for valuable discussions. The work
has been partially supported by Electronic Communication Frontier
Research and Development Grant of the Ministry of Post and
Telecommunications, Japan, and Russian Fund for Basic Research.


\begin{thebibliography}{10}

\bibitem{LevineR}
B.~F. Levine,
\newblock J. Appl. Phys. {\bf 74}, R1 (1993).

\bibitem{ZielBFl59}
A.~{van~der~Ziel},
\newblock {\em Fluctuation Phenomena in Semiconductors},
\newblock Butterworths, London, 1959.

\bibitem{KoganNB96}
S.~Kogan,
\newblock {\em Electronic Noise and Fluctuations in Solids},
\newblock Cambridge University Press, 1996.

\bibitem{LiuNAPL92}
H.~C. Liu,
\newblock Appl. Phys. Lett. {\bf 61}, 2703 (1992).

\bibitem{BeckN93}
W.~A. Beck,
\newblock Appl. Phys. Lett. {\bf 63}, 3589 (1993).

\bibitem{LiuN94}
B.~Xing, H.~C. Liu, P.~H. Wilson, M.~Buchanan, Z.~R. Wasilewski, and J.~G.
  Simmons,
\newblock J. Appl. Phys. {\bf 76}, 1889 (1994).

\bibitem{WangN94}
D.~Wang, G.~Bosman, and S.~S. Li,
\newblock Appl. Phys. Lett. {\bf 65}, 183 (1994).

\bibitem{ChoiAPL94}
K.~K. Choi,
\newblock Appl. Phys. Lett. {\bf 65}, 1266 (1994).

\bibitem{ShadrinN1}
V.~D. Shadrin, V.~V. Mitin, V.~A. Kochelap, and K.~K. Choi,
\newblock J. Appl. Phys. {\bf 77}, 1771 (1995).

\bibitem{LiuFast86}
D.~D. Coon, R.~P.~G. Karunasiri, and H.~C. Liu,
\newblock J. Appl. Phys. {\bf 60}, 2636 (1986).

\bibitem{KingstonB79}
R.~H. Kingston,
\newblock {\em Detection of Optical and Infrared Radiation},
\newblock Springer-Verlag, 1979.

\bibitem{Liu82GHz}
H.~C. Liu, J.~Li, E.~R. Brown, K.~A. McIntosh, K.~B. Nichols, and M.~J. 
Manfra,
\newblock Appl. Phys. Lett. {\bf 67}, 1594 (1995).

\bibitem{LiuSSST91}
H.~C. Liu, M.~Buchanan, G.~C. Aers, and Z.~R. Wasilewski,
\newblock Semicond. Sci. Technol. {\bf 6}, C124 (1991).

\bibitem{RosencherS92}
E.~Rosencher, F.~Luc, P.~Bois, and S.~Delaitre,
\newblock Appl. Phys. Lett. {\bf 61}, 468 (1992).

\bibitem{BandaraS93}
K.~M.~S.~V. Bandara, B.~F. Levine, R.~E. Leibenguth, and M.~T. Asom,
\newblock J. Appl. Phys. {\bf 74}, 1826 (1993).

\bibitem{BandaraST93}
K.~M.~S.~V. Bandara, B.~F. Levine, and M.~T. Asom,
\newblock J. Appl. Phys. {\bf 74}, 346 (1993).

\bibitem{CarboneS97}
A.~Carbone and P.~Mazzetti,
\newblock Appl. Phys. Lett. {\bf 70}, 28 (1997).

\bibitem{Ryzhii_S_JAP95}
V.~Ryzhii and M.~Ershov,
\newblock J. Appl. Phys. {\bf 78}, 1214 (1995).

\bibitem{Ryzhii_S_SST95}
V.~Ryzhii and M.~Ershov,
\newblock Semicond. Sci. Technol. {\bf 10}, 687 (1995).

\bibitem{ErshovIEEE96}
M.~Ershov, V.~Ryzhii, and K.~Saito,
\newblock IEEE Trans. Electron Devices {\bf 43}, 467 (1996).

\bibitem{notation} The sign convention used in this letter
is most simply understood if the electron charge is considered
to be positive (this obviously does not affect the final results),
then the direction of the current coincides with 
the direction of electron transport. 
        Alternatively, one can consider true negative electron
charge, then all currents ($I_e$, $I_w$, etc.) are also negative 
and $e<0$. 

\bibitem{ErshovAPLTr}
M.~Ershov,
\newblock Appl. Phys. Lett. {\bf 69}, 3480 (1996).


\end{thebibliography}

\begin{figure}
\caption{\label{fig:1}
Schematic diagram of the conduction band profile and currents in an SQWIP
(the barrier shapes are arbitrary).
SQWIPs with thermoactivated as well as tunneling transport
can be described by the model used. }
\end{figure}

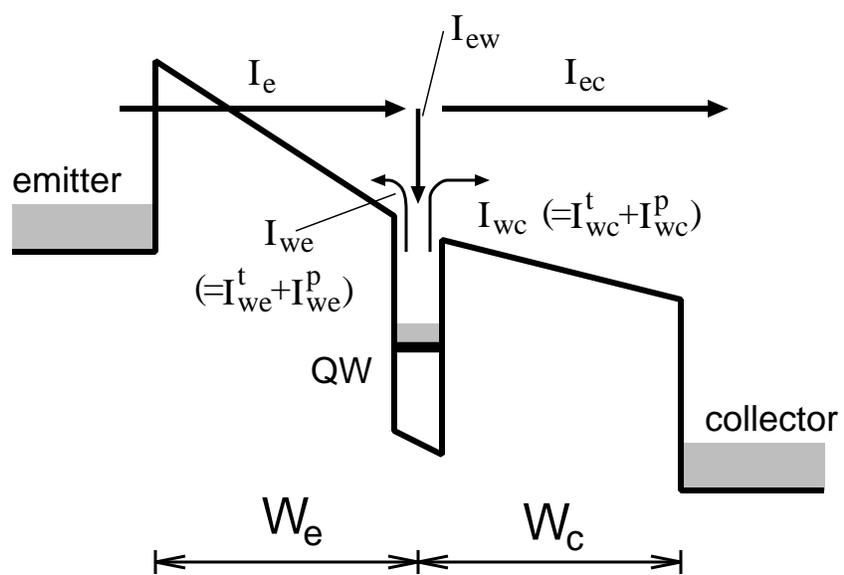
\end{document}